\documentclass[12pt,a4paper,epsf]{article}
\usepackage{graphics}
\usepackage{amssymb,amsmath}
\usepackage[dvips]{lscape,graphicx}
\usepackage{cite}

\newcommand{\be}{\begin{equation}}
\newcommand{\ee}{\end{equation}}

\begin{document}  
\topmargin 0pt
\oddsidemargin=-0.4truecm
\evensidemargin=-0.4truecm
\renewcommand{\thefootnote}{\fnsymbol{footnote}}
\newpage
\setcounter{page}{0}
\begin{titlepage}   
\vspace*{-2.0cm}  
\begin{flushright}
\end{flushright}
\vspace*{0.1cm}
\begin{center}
{\Large \bf Light sterile neutrinos, spin flavour precession and the solar
neutrino experiments
} \\ 
\vspace{0.6cm}
\vspace{0.4cm}

{\large 
C. R. Das\footnote{E-mail: crdas@cftp.ist.utl.pt},
Jo\~{a}o Pulido\footnote{E-mail: pulido@cftp.ist.utl.pt}\\
\vspace{0.15cm}
{  {\small \sl CENTRO DE F\'{I}SICA TE\'{O}RICA DAS PART\'{I}CULAS (CFTP) \\
 Departamento de F\'\i sica, Instituto Superior T\'ecnico \\
Av. Rovisco Pais, P-1049-001 Lisboa, Portugal}\\
}}
\vspace{0.25cm}
and \\
\vspace{0.25cm}
{\large Marco Picariello\footnote{E-mail: Marco.Picariello@le.infn.it}} \\
\vspace{0.15cm}
{\small \sl  I.N.F.N. - Lecce, and
Dipartimento di Fisica, Universit\`a di Lecce\\
Via Arnesano, ex Collegio Fiorini, I-73100 Lecce, Italia}
\end{center}
\vglue 0.6truecm

\begin{abstract}
We generalize to three active flavours a previous two flavour model for the 
resonant spin flavour conversion of solar neutrinos to sterile ones, a mechanism 
which is added to the well known LMA one. The transition magnetic moments
from the muon and tau neutrinos to the sterile play the dominant role in
fixing the amount of active flavour suppression. We also show, through 
numerical integration of the evolution equations, that the data from all solar 
neutrino experiments except Borexino exhibit a clear preference for a sizable 
magnetic field either in the convection zone or in the core and radiation zone. 
This is possibly related to the fact that the data from the first set are 
average ones taken during a period of mostly intense solar activity, whereas 
in contrast Borexino data were taken during a period
of quiet sun. We argue that the solar neutrino experiments are capable of 
tracing the possible modulation of the solar magnetic field. Those monitoring 
the high energy neutrinos, namely the $^8 B$ flux, appear to be sensitive to a 
field modulation either in the convection zone or in the core and radiation 
zone. Those monitoring the low energy fluxes will be sensitive to the second
type of solar field profiles only. In this way Borexino alone may play an 
essential role, since it examines both energy sectors, although experimental 
redundancy from other experiments will be most important.

\end{abstract}

\end{titlepage}   
\renewcommand{\thefootnote}{\arabic{footnote}}
\setcounter{footnote}{0}
\section{Introduction and Motivation} 

Although the effort in solar neutrino investigation has decreased in recent 
years, several intriguing questions in this area remain open. Their 
clarification may lead to a better knowledge of the neutrino intrinsic 
properties, the structure of the inner solar magnetic field, or possibly
both. In fact it is still unclear for example whether the active solar 
neutrino flux varies in time \cite{Sturrock:2008qa,Sturrock:2008jm,
Pandola:2004gz,Sturrock:2004jv} or why the SuperKamiokande 
energy spectrum appears to be flat \cite{Fukuda:2001nj,Fukuda:2002pe}. 
The generally acknowledged LMA solution \cite{Anselmann:1992kc} to the neutrino deficit  
observed by all solar neutrino experiments does not explain these facts, 
while it further predicts an event rate for the Chlorine experiment 
\cite{Cleveland:1998nv} which is 2$\sigma$ above the observed one
\cite{deHolanda:2003tx}. These could be indications of physics beyond LMA.

Based on this motivation and in line with the originally proposed resonant 
spin flavour precession of solar neutrinos \cite{Lim:1987tk,
Akhmedov:1988uk}, we were lead 
to develop a model whereby neutrinos are endowed with a transition magnetic 
moment converting active into sterile ones by virtue of its interaction with 
the magnetic field of the sun \cite{Chauhan:2004sf}. In particular we 
considered a scenario in 
which the apparently time varying event rate of the Gallium experiments 
\cite{Cattadori:2005gq,Gavrin:2007wc}
was viewed in connection with the solar magnetic activity\cite{Chauhan:2006yd}. 
Owing to the uncertainties involved, such data are 
however also consistent with a constant Ga rate, which is the alternative 
conventional view we will consider in the present paper. It is also possible
to generate time variations of the active neutrino flux using the parametric 
resonance for matter density perturbations in the presence of a radiation zone 
magnetic field without resorting to magnetic moments or sterile neutrinos 
\cite{Koutvitsky:2008kd}.

The model expound in refs.\cite{Chauhan:2004sf,Chauhan:2006yd} considered 
two resonances, the LMA one between
two oscillating active neutrinos and the spin flavour precession one determined
by a transition moment between one of the active flavours and the sterile one.
The location of the active-sterile resonance, near the bottom of the solar 
convection zone, was fixed by the corresponding active sterile mass squared 
difference which for this purpose was chosen to be O $(10^{-8}-10^{-9})$ eV$^2$. 
The survival and transition probabilities were calculated using the Landau-Zener 
approximation.

In this paper we extend our previous model with two active flavours and one 
sterile to the more 
realistic case of three active flavours and a sterile. Whereas the Landau Zener 
approximation works well in the LMA resonance, this is not so 
for spin flavour precession, thus we will resort to the numerical integration
of the evolution equations. However the best fit parameter values are obtained
from the Landau Zener approximation, as they are found to coincide in both 
methods. We take several values of $\theta_{13}$ 
in the allowed range for both strong and weak solar fields. The model event 
rates for all solar neutrino experiments are evaluated and confronted with 
the data. Special emphasis is given to the SuperKamiokande energy spectrum 
\cite{Fukuda:2002pe} and the recent $^8B$ energy spectrum from the Borexino 
experiment\cite{Collaboration:2008mr}. We consider two classes of solar 
field profiles, whose field strengths peak in the solar core (Wood-Saxon 
potential type) and in the lower convection zone. Together these span 
a large number of possibilities.

The paper is organized as follows: in section 2 the derivation of the (4$\times$4) 
Hamiltonian in the mass eigenstate basis is presented. In subsection 3.1
the already known fact that the survival probability is a decreasing 
function of $\theta_{13}$ is shown to follow from a simple argument. In
subsection 3.2 the evaluation of the event rates and spectra is presented.
The results are discussed in subsection 3.3. Finally in 
section 4 our conclusions are drawn. Our numerical calculations 
are based on the updated central values \cite{Fogli:2008ig} for 
$\Delta m^2_{21}$, $\theta_{12}$, $\theta_{23}$, $\Delta m^2_{32}$ and we
use a neutrino transition moment between flavour states not larger than
$\mu_{\nu}=1.4\times 10^{-12}\mu_B$. As for $\theta_{13}$ we chose to investigate 
three cases: $\theta_{13}=0$, 0.1 and the central value, 0.13. The fits to all 
data, including rates and spectra (except for Borexino) improve once the 
magnetic field is introduced. As regards Borexino, the fit worsens in this 
case. This may be connected to the fact that all former experiments report 
time averaged data taken during times of more or less intense solar magnetic 
activity, whereas Borexino data were taken during a period of minimum 
activity. In contrast, solar data alone show no clear preference for a 
vanishing or sizable $\theta_{13}$.

\section{The Hamiltonian}

The (4$\times$4) Hamiltonian involving one sterile neutrino and three active ones 
will be expressed in the mass eigenstate basis. This is related to the flavour 
basis by
\be
\left(\begin{array}{c}\nu_{S}\\ \nu_{e} \\ \nu_{\mu} \\
\nu_{\tau} \end{array}\right)=U^{PMNS}(4{\rm x}4)\left(\begin{array}{c}\nu_{0}\\
\nu_{1} \\ \nu_{2} \\ \nu_{3} \end{array}\right)
\ee
where $U^{PMNS}$(4$\times$4) is the straightforward (4$\times$4) extension of the 
usual leptonic mixing matrix \cite{Amsler:2008zzb}. As before \cite{Chauhan:2004sf,
Chauhan:2006yd} no vacuum mixing between the 
sterile and any of the active flavours is assumed, so that the free propagating 
term of the Hamiltonian is in the mass basis 
\be
(H_0)_M =\left(\begin{array}{cccc} E_0 & & & \\ & E_1 & &  \\ & & E_2 & \\
& & & E_3 \end{array}\right)
\ee
and the matter (interaction) term is in the flavour basis
\be
(H_I)_W =\left(\begin{array}{cccc} 0 & \mu_{es} B & \mu_{\mu s} B & \mu_{\tau s} B 
\\ \mu_{es} B & V_c+V_n & 0 & 0 \\
\mu_{\mu s} B & 0 & V_n & 0 \\ \mu_{\tau s} B & 0 & 0 & V_n\\ \end{array}\right).
\ee
Here $\mu_{(e,\mu,\tau) s}$ are the transition magnetic moments between the 
active flavours and the sterile one and $B$ is the magnetic field profile. 
The quantities $V_c$, $V_n$ are the refraction indexes for 
charged and neutral currents, namely $V_c=G_F \sqrt{2}N_e$, $V_n=-(G_F /\sqrt{2}) 
N_{n}$ with $N_e$ (electron density) and $N_n$ (neutron density).
Given eq.(1), the mass and flavour Hamiltonian 
representations are related by 
\be
H_{M}=U^{\dagger PMNS}H_W U^{PMNS}
\ee 
so 
that the full Hamiltonian in the mass basis is 
\be
H_{M}=\left(\begin{array}{cccc} E_0 & & & \\ & E_1 & &  \\ & & E_2 & \\
& & & E_3 \end{array}\right)+U^{\dagger PMNS}\left(\begin{array}{cccc}
0 & \mu_{es} B & \mu_{\mu s} B & \mu_{\tau s} B \\ \mu_{es} B & V_c+V_n & 0 & 0 \\
\mu_{\mu s} B & 0 & V_n & 0 \\ \mu_{\tau s} B & 0 & 0 & V_n\\ \end{array}\right) U^{PMNS}.
\ee
Subtracting $E_1$ and denoting by $\tilde \mu_{1,2,3}$ the transition magnetic
moment between mass eigenstates 0 and 1,2,3 respectively, we have
\be
H_M=\left(\begin{array}{cccc} \frac{\Delta m^2_{01}}{2E} & {\tilde \mu_1}B 
& {\tilde \mu_2}B & {\tilde \mu_3}B\\
\tilde \mu_1 B & 0 & 0 & 0 \\ \tilde \mu_2 B & 0 & \frac{\Delta m^2_{21}}{2E} & 0 \\
\tilde \mu_3 B& 0 & 0 & \frac{\Delta m^2_{31}}{2E} \end{array}\right)+U^{\dagger PMNS}
\left(\begin{array}{cccc}
0 & 0 & 0 & 0 \\ 0 & V_c+V_n & 0 & 0 \\
0 & 0 & V_n & 0 \\ 0 & 0 & 0 & V_n\\ \end{array}\right) U^{PMNS}.
\ee
Straightforward matrix algebra leads now to 
\be
H_M=\left(\begin{array}{cccc} \frac{\Delta m^2_{01}}{2E} & \tilde \mu_1 B 
& \tilde \mu_2 B & \tilde \mu_3 B\\
\tilde \mu_1 B & V_n+V_c u^2_{e_1} & V_c u_{e_1}u_{e_2} & V_c u_{e_1}u_{e_3} \\
\tilde \mu_2 B& V_c u_{e_1}u_{e_2} & \frac{\Delta m^2_{21}}{2E}+V_n+V_c u^2_{e_2}
& V_c u_{e_2}u_{e_3} \\ \tilde \mu_3 B& V_c u_{e_1}u_{e_3} & V_c u_{e_2}u_{e_3} &
\frac{\Delta m^2_{31}}{2E}+V_n+V_c u^2_{e_3} \end{array}\right)
\ee
where $u_{e_i}$ denotes the first row entries of the (3$\times$3) $U^{PMNS}$ matrix.
In the following we assume vanishing phases. Eq. (7) is the mass basis
Hamiltonian that we use throughout.

\section{Probability and Rates}

\subsection{3 Flavour probability and $\theta_{13}$}
A simple argument shows that the electron neutrino survival probability for 
three active flavours \cite{Fogli:1999zg,Balantekin:2003dc} 
\be
P_{3\times3} (\nu_e \rightarrow \nu_e)=\cos^{4}\theta_{13}P_{2\times2} (
\nu_e \rightarrow \nu_e (N_e\rightarrow N_e \cos^{2}\theta_{13}))+\sin^{4}\theta_{13}
\ee
where extra terms $O(10^{-3})$ or smaller have been 
neglected \cite{Balantekin:2003dc}, is a decreasing function of $\theta_{13}$. In 
eq.(8) the (2$\times$2) probability with the replacement $N_e\rightarrow N_e \cos^{2}
\theta_{13}$ is given by \cite{Balantekin:2003dc} 
\be
P_{2\times2}=\frac{1}{2}+\frac{1}{2}\cos2\theta_{12}\frac{-\phi(x)}
{\sqrt{(\frac{\Delta m^2_{21}}{4E}\sin 2\theta_{12})^2+\phi^2}}
\ee
where $x$ is the fractional solar radius and  
\be
\phi=\frac{G_F}{\sqrt{2}}N_e \cos^{2} \theta_{13}-\frac{\Delta m^{2}_{21}}{4E}
\cos2\theta_{12}.
\ee
Straightforward calculations show that the derivative of $P_{3\times3} 
(\nu_e \rightarrow \nu_e)$ with respect to $\theta_{13}$ is negative for all 
solar neutrino energies if
\be
P_{2\times2}-\frac{G_F N_e}{4\sqrt{2}}~\cos2\theta_{12}~\cos^{2} \theta_{13}~\frac{
(\frac{\Delta m^2_{21}}{4E}\sin2\theta_{12})^2}{[(\frac{\Delta m^2_{21}}{4E}
\sin2\theta_{12})^2+\phi^2]^{\frac{3}{2}}}-\tan^{2}\theta_{13}>0.
\ee
In fig.1 we plot the quantity on the left hand side of this inequality as a function 
of the neutrino production point denoted by its fractional radius $x$, for energies 
$E=0.1,~8,~18.8$ MeV. $N_e$ is evaluated from the data on $\rho$ and $X_H$ given in 
\cite{Bahcallhomepage} with
\be
N_e=\frac{\rho}{m_p}\frac{1+X_H}{2}
\ee
where $\rho$, $m_p$ and $X_H$ are the density, the proton mass and the hydrogen mass 
fraction. All neutrino parameters including $\theta_{13}$ were fixed at their best 
fit central values 
\cite{Fogli:2008ig}. The minimum of the quantity (11) is clearly seen in fig.1 for 
the three energy values considered and the central value $\theta_{13}=0.13$ claimed 
in ref.\cite{Fogli:2008ig}. For $x=r/R_{\odot}$, this minimum reaches zero at  
$x\simeq 0.2$ as the energy is increased up to $E_{max}=E_{max}(hep)=18.8$ MeV 
and $\theta_{13}$ up to 0.47 which is much above $3\sigma$ from its central value 
\footnote{It has been recently pointed out \cite{Schwetz:2008er,Maltoni:2008ka} 
that the hint for a 
non-vanishing $\theta_{13}$ is at most a 1$\sigma$ effect on the basis of present 
data.}. Thus we can conclude that for all experimentally allowed values 
of the physical quantities involved the condition (11) is satisfied and $P_{3\times3} 
(\nu_e \rightarrow \nu_e)$ is a decreasing function of $\theta_{13}$.

\subsection{Field profiles and rates}

We base our numerical calculations for the rates on the standard solar model with 
high metalicity, BPS08(GS) \cite{PenaGaray:2008qe}. As far as the solar field is
concerned, solar physics provides very limited knowledge on its magnitude and shape.
For instance in ref. \cite{Boruta} upper bounds of 0.5-1.5 G and 30 G are quoted in
the bottom of the convection zone and in the core respectively while in ref. 
\cite{Kitchatinov} an argument is presented favouring an upper bound of 600 G in the
radiation zone. On the other hand, the authors of ref. \cite{Antia:2000pu} estimate 
in the bottom of the convection zone an upper limit of 300 kG and in the 
mid-radiation zone and solar centre a magnetic field strength of 0.7 MG and 7 MG 
respectively \cite{Antia:2008}. 

Given the above mentioned uncertainties we consider the two following plausible 
profiles which are approximately complementary to each other (see also fig.2)

{\it Profile 1}
\be
B=\frac{B_0}{\cosh[6(x-0.71)]} \quad 0<x<0.71
\ee
\be
B=\frac{B_0}{\cosh[15(x-0.71)]} \quad 0.71<x<1
\ee

{\it Profile 2}
\be
B=\frac{B_0}{1+\exp[10(2x-1)]} \quad 0<x<1,
\ee
Profile 1 has a peak $B_0$ at the bottom of the convection zone, for fractional 
solar radius $x\simeq 0.71$, its physical motivation being the 
large gradient of angular velocity over this range \cite{Antia:2008dn}. It should 
not exceed 300 kG at this depth and 20 kG at 4-5\% depth, hence its fast decrease 
along the convection zone \cite{Antia:2000pu}. Profile 2 is of the Wood-Saxon type, 
being maximal at the solar centre. In this case the peak field $B_0$ could be as 
large as a few MG \cite{Rashba:2006jp}.

The survival probability was obtained from the numerical integration 
of the Schr\"{o}dinger like equation with Hamiltonian (7) using the Runge-Kutta 
method. All event rates (total and 
spectral) were evaluated as described in refs. \cite{Pulido:1999xp,Pulido:2001bd}.  
The expression for the $^8 B$ spectral rate as applied to the SuperKamiokande and 
Borexino experiments with three active neutrinos is now
\be
R^{th}_{SK,Bor}(E_e)=
\frac{\displaystyle\int_{m_e}^{{E'_e}_{max}}dE'\!\!_ef_{SK,Bor}(E'_e,E_e)
\int_{E_m}^{E_M}dE\phi(E)\left[P_{ee}(E)\frac{d\sigma_{e}}{dT'}+
\left(P_{e\mu}(E)+P_{e\tau}(E)\right)\frac{d\sigma_{\mu,\tau}}{dT'}\right]}
{\displaystyle\int_{m_e}^{{E'_e}_{max}}dE'\!\!_e
f_{SK,Bor}(E'_e,E_e)\int_{E_m}^{E_M}dE\phi(E)\frac{d\sigma_{e}}{dT'}}
\ee
where $\sigma_{e}$ is the charged and neutral current cross section and
$\sigma_{\mu,\tau}$ is the neutral current one. 
The energy resolution functions are given in \cite{Fukuda} and 
\cite{Montanino:2008hu} and the threshold energies $E_e=5$ MeV, $E_e=2.8$ MeV for
SuperKamiokande and Borexino respectively. For the statistical analysis of all
solar data (except Borexino) we used the standard $\chi^2$ definition
\cite{Pulido:1999xp,Pulido:2001bd}
\be
\chi^2=\sum_{j_{1},j_{2}}({R}^{th}_{j_{1}}-{R_{j_{1}}}^{\exp})\left[{\sigma^2}
(tot)\right]^{-1}_{j_{1}j_{2}}({R}^{th}_{j_{2}}-{R_{j_{2}}}^{\exp})
\ee
where indices $j_{1},j_{2}$ run over solar neutrino experiments and the error matrix
includes the cross section, the astrophysical and the experimental uncertainties.

The rates we obtained were confronted with the data from the Cl experiment 
\cite{Cleveland:1998nv}, the Ga ones \cite{Cattadori:2005gq}, the SuperKamiokande
spectrum \cite{Fukuda:2002pe}, the SNO rates and spectra \cite{Aharmim:2005gt},
the Borexino spectra for $^8 B$ neutrinos \cite{Collaboration:2008mr} and for 
the remaining fluxes \cite{Arpesella:2007xf} \footnote{We recall that, as 
mentioned in the introduction, we neglect any possible time variation in the Ga 
rate and perform fits to its average value, see ref. \cite{Cattadori:2005gq}.}. 
Starting with profile 1 (fig.2) which peaks at the bottom of the convection zone, 
we considered the case of a relatively strong field and a vanishing one, and 
likewise for profile 2 (fig.2). Numerical results are insensitive to field values 
below $50$ kG. In order to
provide a feeling of the rates variation with $\theta_{13}$, we also run this 
parameter from zero to 0.13. The values of the active sterile mass squared 
difference, which determines the location of the spin flavour resonance, 
were $\Delta m^2_{01}=m_{0}^2-m_{1}^2=1.25\times 10^{-7}$ eV$^2$ and 
$\Delta m^2_{01}=2.7\times 10^{-6}$ eV$^2$ obtained by fitting with profile 1 
and profile 2 respectively, using the Landau Zener approximation for the spin
flavour precession resonance. For the $^8 B$ flux normalization 
we used $f_B=0.95$ \cite{Collaboration:2008mr} and the remaining neutrino 
parameters were \cite{Fogli:2008ig}
\be
\Delta m^2_{21}=7.67\times 10^{-5}{\rm eV}^2,~\Delta m^2_{23}=2.39\times 10^{-3}{\rm eV}^2
,~\sin\theta_{12}=0.559,~\sin\theta_{23}=0.683.
\ee
Since, as explained below, our predictions refer to the average magnetic
activity in the solar cycle, they should be evaluated for an intermediate field
strength. Therefore we take $B_0=140$ kG for profile 1, while for profile 2
the peak value will be obtained from fitting. Regarding the mass basis magnetic 
moments $\tilde \mu_{1,2,3}$, we found that the numerical 
results are independent of which moment is chosen to be the largest. For the 
assumed peak field strength, the fits require this largest value to be 
$2\times 10^{-12}\mu_B$. To this end the two flavour (transition) moments 
$\mu_{(\mu,\tau)s}$ must be of order $1.4\times 10^{-12}\mu_B$ with $\mu_{es}$
equal or arbitrarily smaller. A value $1.7\times 10^{-12}\mu_B$ is also possible
for the largest of the mass basis magnetic moments\footnote{This would however
require $B_0\simeq (160-170)$ kG.}, provided either $\mu_{\mu s}$
or $\mu_{\tau s}$ is as large as $1.4\times 10^{-12}\mu_B$ with the other two
no smaller than $1.0\times 10^{-12}\mu_B$. In other words, $\mu_{(\mu,\tau)s}$
are the dominant moments that fix the amount of the active neutrino flavour
suppression. In the following we will consider $\mu_{(\mu,\tau)s}
\simeq 1.4\times 10^{-12}\mu_B$ with $\mu_{es}$ arbitrarily equal or smaller,
thus ensuring that the largest of the $\tilde \mu$'s is $\simeq 2\times 10^{-12}
\mu_B$.

\subsection{Discussion}

Starting with profile 1, the results 
for all solar neutrino experiments except Borexino, which is discussed below, 
are displayed in table I (total rates and fluxes) and fig.3 (SuperKamiokande 
spectral rate). They show a clear preference for a sizable magnetic field. 
We note that not only the flatness of the spectrum is enhanced thus providing 
a better fit to the data (see fig.3), but also the total rate predictions for 
the Cl, SK and SNO experiments strongly improve (table I). A strictly constant 
spectrum could on the other hand be obtained by varying the solar parameters 
$\Delta m^2_{21}$ and $\sin\theta_{12}$ within a 2$\sigma$ range. As for the Ga 
rate, vanishing and sizable fields are equivalent, as both classes of 
predictions lie within $1\sigma$ of the central value \cite{Cattadori:2005gq}. 
Moreover as to the magnitude of $\theta_{13}$, the predictions do not show 
any clear preference. Altogether the results are more sensitive to changes
in the solar parameters than in the atmospheric ones.

\begin{center}
\begin{tabular}{cc|cccccccccc} \\ \hline \hline
$B_0(kG)$ & $\sin \theta_{13}$ & Ga & Cl & SK & $\rm{SNO_{NC}}$ & $\rm{SNO_{CC}}$ & $\rm{SNO_{ES}}$ &
$\!\!\chi^2_{rates}\!\!$ & $\chi^2_{{SK}_{sp}}$ & $\chi^2_{SNO}$ & $\chi^2_{gl}$\\ \hline
 & 0  & 67.2 & 2.99 & 2.51 & 5.62 & 1.90 & 2.49 & 0.07 & 42.7 & 57.2 & 99.9 \\
0 & 0.1 & 66.0 & 2.94 & 2.49 & 5.62 & 1.87 & 2.46 & 0.30 & 42.1 & 55.2 & 97.6 \\ 
 & 0.13 & 65.0 & 2.90 & 2.46 & 5.62 & 1.84 & 2.44 & 0.62 & 41.7 & 53.7 & 96.0 \\
\hline
  & 0  & 66.4 & 2.82 & 2.32 & 5.37 & 1.76 & 2.31 & 0.20 & 37.6 & 46.0 & 83.8 \\
140  & 0.1  & 65.3 & 2.77 & 2.29 & 5.37 & 1.73 & 2.28 & 0.53 & 37.9 & 44.9 & 83.3 \\
  & 0.13  & 64.3 & 2.72 & 2.27 & 5.37 & 1.70 & 2.25 & 0.95 & 38.4 & 44.1 & 83.4 \\
\hline
\end{tabular}
\end{center}
{\it{Table I - Peak field values (profile 1), $\sin \theta_{13}$, total rates 
(in SNU for Ga and Cl experiments, in $10^6 cm^{-2}s^{-1}$ for SK and SNO), and 
the corresponding $\chi^{2}$'s. The total number of degrees of freedom is 82 = 84 
experiments (Ga + Cl + 44 SK + 38 SNO data points) - 2 parameters, (see ref.
\cite{Chauhan:2006yd}). It is seen that for a sizable field ($B_0=140$ kG) all 
fits improve.}}

\vspace{0.6cm}

The Borexino spectral rate for $^8 B$ is shown in fig.4. The top curve is the 
central value of the ratio between the best fit recoil spectrum due to oscillated 
neutrinos and the spectrum due to non-oscillated ones evaluated from fig.3 of ref.
\cite{Collaboration:2008mr}. The next group of three curves represents the 
predicted spectra for vanishing field with $\theta_{13}=0,0.1,0.13$ from top to
bottom respectively, and the bottom three curves represent the same but for a 
field ($B_0=140$ kG). The result of a $\chi^2$ analysis with 4 degrees of 
freedom for vanishing field and profile 1 with $B_0=140$ kG is presented in the 
first three columns of table II. Owing to the magnitude of the experimental 
errors, it is not possible to conclude whether the data shows any preference 
for a vanishing or a finite $\theta_{13}$, although a sizable magnetic field 
appears to be disfavoured. From fig.4 and table II our predictions look more 
sensitive to the magnetic field variation than to the $\theta_{13}$ one within 
their respective allowed ranges. An improved significance can be obtained if 
Borexino are able to reduce their errors to 1/3 of the present ones (see the
fourth column of table II).

\begin{center}
\begin{tabular}{cccc} \hline \hline
$B_0(kG)$ &  $sin\theta_{13}$ & $\chi^2$ & $\Delta \chi^2$ \\ \hline
  0   &       0           &   4.55  &      0   \\
  0   &      0.1        &   4.55  &      0   \\
  0   &      0.13         &   4.56  &      0   \\  
140   &       0           &   4.93  &     2.4  \\
140   &      0.1        &   4.98  &     2.5  \\
140   &      0.13         &   5.03  &     2.6  \\
\hline
\end{tabular}
\end{center}
{\it{Table II - The result of our $\chi^2$ analysis for Borexino (profile 1): 
from the first three columns it is seen that no conclusion can be drawn as for 
the magnitude of $sin\theta_{13}$ and that the significance is too low
for the data to favour a vanishing field. The last column shows the $\chi^2$ 
variation if the experimental error were reduced to 1/3, so 
a vanishing field would clearly be favoured. The result would still 
in this case be inconclusive regarding the size of $sin\theta_{13}$.}}

\vspace{0.6cm}

From the data and the model predictions presented, it is unclear whether Borexino 
can favour either a negligible or a sizable solar magnetic field. The available 
data from the remaining experiments appear to favour on the other hand
a sizable field, possibly connected to a more intense solar 
activity in the convection zone. Hence it might be appropriate to examine the 
period in which the data were taken. In particular the SuperKamiokande spectrum 
refers to the period from May 31, 1996 to July 15, 2001 during which the average 
sunspot number was 65. On the other hand the Borexino $^8 B$ spectrum refers to 
a data taking period from July 15, 2007 to June 21, 2008 with average sunspot 
number 4 \cite{Solar}. In most of the former period the solar magnetic activity 
increased and reached an 11-year peak in the Summer of 2000, whereas in the 
latter the activity was continuously at its minimum. Therefore in the light of 
the present model, one expects the present Borexino spectrum for $^8 B$ to 
coincide with the LMA prediction and the SuperKamiokande one to reflect an 
active sun.

We have also tested the model for the remaining fluxes observed in the 
first Borexino phase \cite{Arpesella:2007xf}, including in addition the $pp$ 
and $pep$ neutrinos. In this case $E\leq 1.7$ MeV for all fluxes,
so that with $\Delta m^2_{01}=1.25\times 10^{-7}$ eV$^2$ all neutrino resonances 
lie below $x=0.5$ where the magnetic field strength is $B<\frac{1}{2}B_0$
(see fig.2). Furthermore since the matter density is larger in this range, 
the variation of the field besides being smaller, becomes much less important. 
The maximum variation that is reflected in the event rates is no greater than
1\%, hence, given the order of magnitude of the experimental errors which is
about 25\% \cite{Arpesella:2007xf}, it cannot be expected to be seen in this 
case for any flux. We may therefore conclude that it is of prime importance 
that Borexino will continue monitoring both the low energy and the $^8 B$ flux 
during the present increasing solar activity period. 

While it is generally accepted that the sunspot activity is interrelated with the 
possible modulation of the convection zone magnetic field, no connection
appears to exist between such varying activity and the radiation and core 
magnetic field. There is however a recent claim in the literature 
\cite{Sturrock:2008jm} suggesting the existence of an inner tachocline separating 
the core from the radiation zone and an inner dynamo producing a strong magnetic 
field and a second solar cycle. Independently of the fact that our knowledge on
this matter that can be obtained from solar physics is very limited at present, 
it will be shown in the following that the solar neutrino data are consistent 
with a varying field in these inner regions.

Referring to profile 2 [eq.(15) and fig.2] and in order that a possible time 
modulation may be detected, all resonances 
from active to sterile must be located deeper inside the sun with relation to 
profile 1, so that the shift from a weak to a strong
field or vice versa is reflected in the intensity of the neutrino flux. In
this way the best fit to the data was obtained for $\Delta m^2_{10}=
2.7\times10^{-6}$ eV$^2$ and $B_0=0.75$ MG \footnote{For instance with $\Delta m^2_{10}=
1.25\times10^{-7}$ eV$^2$, the neutrinos with energy $E=5$ MeV have their resonance at
$x=0.64$ whereas for $\Delta m^2_{10}=2.7\times10^{-6}$ eV$^2$ this resonance  
moves to $x=0.34$.} with the remaining neutrino parameters as in eq.(18). 
The results for the total rates and fluxes are now shown in table III 
and those for the SuperKamiokande and $^8 B$ Borexino spectra in figs.5 and 6. 

\begin{center}
\begin{tabular}{cc|cccccccccc} \\ \hline \hline
$B_0(MG)$ & $\sin \theta_{13}$ & Ga & Cl & SK & $\rm{SNO_{NC}}$ & $\rm{SNO_{CC}}$ & $\rm{SNO_{ES}}$ &
$\!\!\chi^2_{rates}\!\!$ & $\chi^2_{{SK}_{sp}}$ & $\chi^2_{SNO}$ & $\chi^2_{gl}$\\ \hline
  & 0  & 64.7 & 2.75 & 2.32 & 5.38 & 1.76 & 2.32 & 0.76 & 38.0 & 46.1 & 84.8 \\
0.75  & 0.1  & 63.6 & 2.70 & 2.30 & 5.38 & 1.73 & 2.29 & 1.32 & 38.4 & 45.0 & 84.7 \\
  & 0.13  & 62.6 & 2.66 & 2.28 & 5.38 & 1.70 & 2.26 & 1.92 & 38.8 & 44.2 & 84.9 \\
\hline
\end{tabular}
\end{center}
{\it{Table III - Same as table I for profile 2 where the vanishing field case is
omitted. As for profile 1, with a sizable field ($B_0=0.75$ MG) all fits improve
with relation to the vanishing field (compare with table I).}}

\vspace{0.6cm}

Again, as for profile 1, the SuperKamiokande data show a clear preference for a 
large field (fig.5) and the quality of the fits is the same for both profiles. 
As can be seen from a comparison between figs.3 and 5 or figs.4 and 6, the spectra 
for a sizable peak field are much alike and it will be very hard to experimentally 
distinguish between them in this way. The actual difference can be explicitly seen 
in fig.7 where the Borexino spectrum for both profiles is shown for $\theta_{13}=0$ 
with the remaining parameters as in tables I and III. Whereas for profile 1 the 
spectrum presents a shallow minimum around $E=8$ MeV, it decreases monotonically 
in the case of profile 2. We have also seen that in the case of profile 2 the 
results are less stable, in the sense that small variations either in the neutrino 
parameters $\Delta m^2_{21}$ and $\theta_{12}$ or $B_0$ lead to larger variations in 
the rates and fits. Similarly to profile 1, the results for a $\chi^2$ analysis for 
profile 2 are shown in table IV and the case for a vanishing or a sizable field is 
again inconclusive.
 
\begin{center}
\begin{tabular}{cccc} \hline \hline
$B_0(MG)$ &  $sin\theta_{13}$ & $\chi^2$ & $\Delta \chi^2$ \\ \hline
0.75  &       0           &   4.93  &     1.9  \\
0.75  &      0.1        &   4.89  &     1.8  \\
0.75  &      0.13         &   4.84  &     1.7  \\
\hline
\end{tabular}
\end{center}
{\it{Table IV - The same as table II for profile 2. The fourth column indicates
the $\chi^2$ variations relative to its values for vanishing field  
given in the first three rows of table II.}}

\vspace{0.6cm}

Finally for profile 2 we have calculated the rates corresponding to the remaining 
fluxes which were observed in the first Borexino phase \cite{Arpesella:2007xf}.
For all these neutrinos, with $E\leq 1.7$ MeV, the corresponding resonances lie 
in the range $x<0.23$ where the field is close to maximal. Thus the event 
rate modulation is expected to be much stronger than with profile 1. It is now
approximately 9\% which we believe to be within reach of the Borexino experiment
in the future. The results are shown in fig.8 where we chose to represent the 
modulation of the $^7 Be$, $^{15} O$ and $^{13} N$ fluxes. Owing to the magnitude
of the errors involved ($\simeq$ 25\%), the $\chi^2$, of order 38 (51 d.o.f.), is 
extremely flat for both profiles and it is hard to distinguish any preference of 
the data at all either for profile 1 or 2. 
So it is also of prime importance to keep Borexino monitoring the low energy 
neutrino fluxes too, namely $^7 Be$ and CNO as in the first phase. Therefore 
solar neutrino experiments hold the potential to clearly trace a field modulation 
inside the sun and moreover possibly to distinguish whether this modulation 
occurs in the convection zone or deeper in the radiation zone and core.

\section{Summary and Conclusions}

We have developed a model with three active neutrino flavours communicating to 
a sterile one in matter with magnetic field through transition magnetic 
moments. Its motivation is to provide better fits to the solar neutrino data 
than the LMA ones, 
in particular a flat SuperKamiokande spectrum and a better prediction 
for the Chlorine rate, while keeping accurate predictions for all other rates 
including the recent $^8 B$ Borexino spectrum. We investigated two magnetic
solar field profiles, one which peaks at the bottom of the convection zone
and another at the solar centre. These represent two classes of plausible
possibilities which somehow complement each other. 

The starting point was the derivation of the (4$\times$4) Hamiltonian followed by
a simple and general argument showing that the survival probability is a
decreasing function of the still unknown mixing angle $\theta_{13}$. This
fact, which is reflected more strongly in the charged current data, leads
to a parallel shift of the spectral event rates. This shift is however not
enough to distinguish a clear preference of the data between a vanishing or
a sizable $\theta_{13}$.

We found that among the
three transition moments, the ones connecting $\nu_{\mu}$ and $\nu_{\tau}$
to the sterile are the dominant ones that fix the amount of active flavour
suppression. They are both required to be of order $1.4\times 10^{-12}\mu_B$,
while $\nu_{es}$ may be equal or arbitrarily smaller. Alternatively either
$\nu_{\mu}$ or $\mu_{\tau}$ separately may be of order $1.4\times 10^{-12}
\mu_B$ with the remaining two of order $1.0\times 10^{-12}\mu_B$ in which case
a slightly stronger field is required.

On the other hand it was found that all experimental data, with the exclusion 
of the Borexino ones, favour a relatively large magnetic field of either class. 
To this end it is important to realize that the former data are average ones and 
refer to extended periods. In particular the SuperKamiokande spectrum refers
to a period when the average solar magnetic activity was relatively intense, 
and hence it is sensible to expect it to be flat in a way that it reflects a 
large field in accordance with the model predictions. As regards the Borexino 
spectrum in a similar energy range, it is 
not possible at present to conclude whether it favours the LMA spectrum or 
the LMA one with spin flavour precession, as the data errors are too large.

Whereas the neutrino fluxes observed in the first Borexino phase were found
to be insensitive to field modulations in the convection zone (profile 1), 
this is not so if the field is concentrated in the core and radiation
zone (profile 2). The event rate modulation expected in this case is of the 
same magnitude as the one expected for the $^8 B$ flux with any of the profiles.
Hence we believe it extremely important to keep Borexino taking data for all
neutrino fluxes during at least the first half of the present solar cycle 
expected to peak in 2011 or 2012.

Our claim is not that there is evidence of variability of the solar field
profile in the convection zone or equivalently in the core and radiative zone,
but rather that the neutrino data are consistent with the possibility of either 
phenomenon.

Our results concerning field profiles and data variability are qualitatively 
summarized in table V: a magnetic field
concentrated around the bottom of the convection zone like profile 1 can only
show its modulation through an experiment monitoring the high energy $^8 B$
flux, whereas a field concentrated in the core and radiation zone like 
profile 2 can be detected by experiments monitoring either the high energy 
$^8 B$ or the low energy fluxes.

To conclude, solar neutrino experiments may hold a
non-negligible potential to ascertain whether there is a varying magnetic field
inside the sun possibly connected to solar activity, a fact which otherwise
may be very difficult to establish on the basis of solar physics alone. Moreover
we have shown that it is also possible to trace whether this varying field is
lying mostly at the bottom of the convection zone or deeper in the core and
radiation zone.

\begin{center}
\begin{tabular}{ccc} \\ \hline \hline
Varying field & $^8 B$ flux & Others  \\ \hline
Profile 1 (CZ) & Yes & No  \\
Profile 2 (WS) & Yes & Yes  \\
\hline
\end{tabular}
\end{center}
{\it{Table V - The possibility for detecting through solar neutrino experiments
the magnetic fields concentrated either in the convection zone (profile 1) or in 
the core and radiation zone (profile 2).}}

\vspace{1cm}
\noindent {\Large \bf Acknowledgments}
\vspace{0.5cm}

{\em C. R. Das gratefully acknowledges a scholarship from Funda\c{c}\~{a}o para
a Ci\^{e}ncia e Tecnologia ref. SFRH/BPD/41091/2007. One of us (M.P.) is grateful
to D. Montanino for useful discussions.}


\newpage

\begin{figure}
\centering
\hspace*{-1.5cm}
\includegraphics[height=120mm,keepaspectratio=true,angle=0]{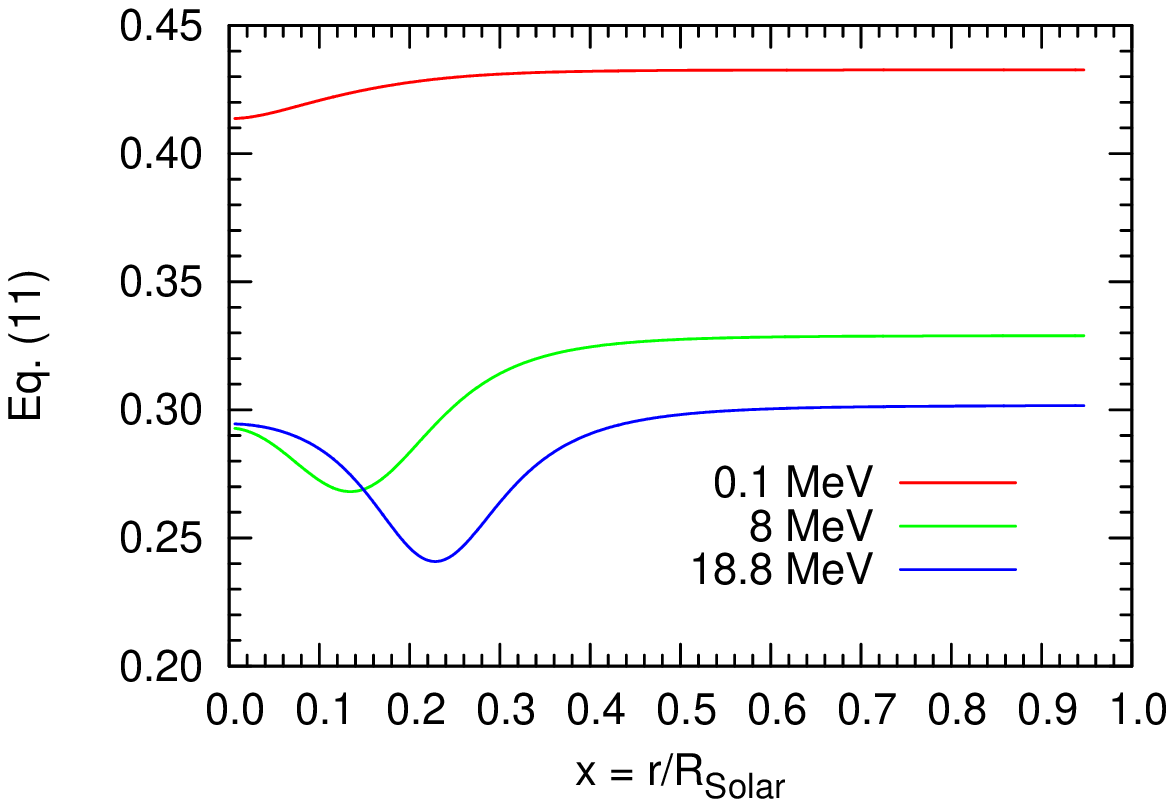}
\caption{ \it
The three lines represent expression (11) as a function 
of the neutrino production point for neutrino energies 0.1 MeV, 8 MeV and 18.8 MeV
with $\sin\theta_{13}=0.13$. For all experimentally allowed values of $\theta_{13}$
the quantity plotted in this graph is positive implying that the survival
probability is always a decreasing function of $\theta_{13}$.}
\label{fig1}
\end{figure}

\begin{figure}
\centering
\hspace*{-1.5cm}
\includegraphics[height=120mm,keepaspectratio=true,angle=0]{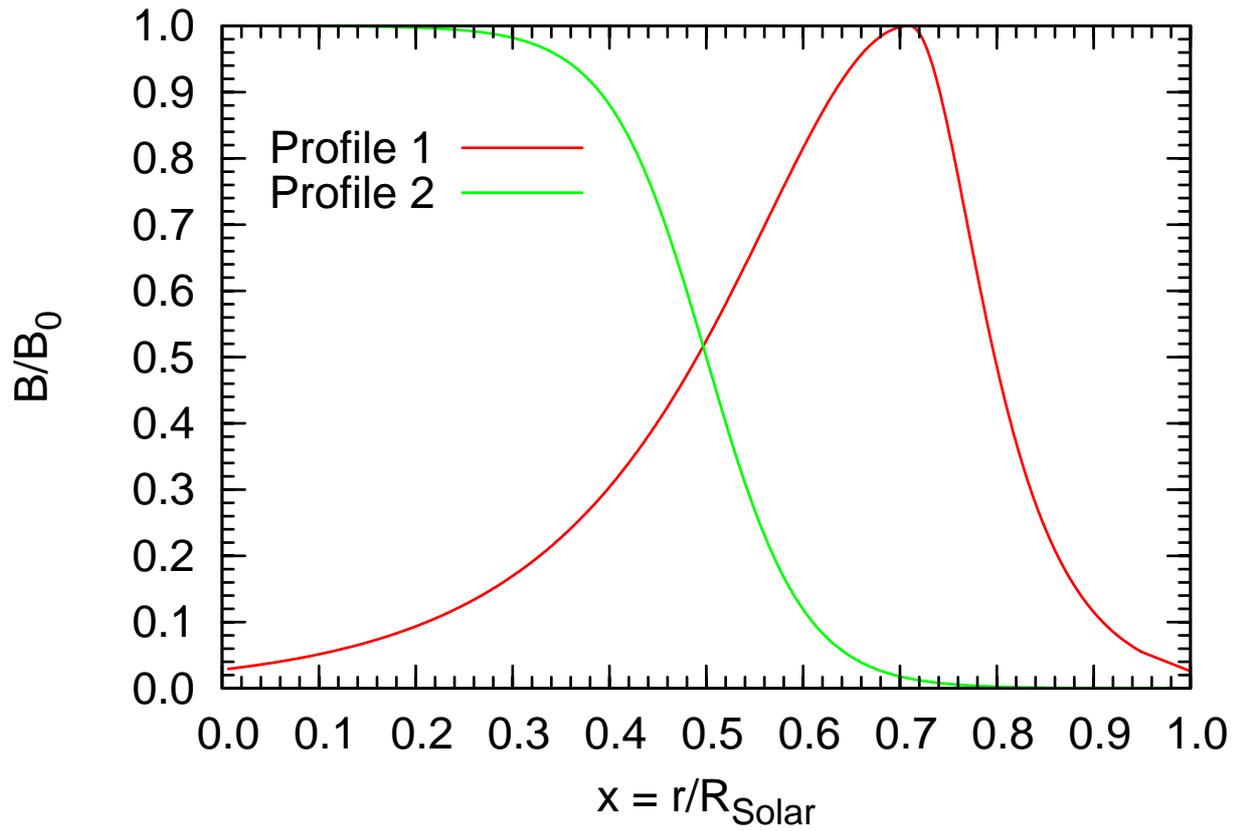}
\caption{ \it The two solar field profiles given by eqs.(13), (14) 
and (15) normalized to their peak field
values and expressed as a function of the fractional solar radius.}
\label{fig2}
\end{figure}

\begin{figure}
\centering
\hspace*{-1.5cm}
\includegraphics[height=120mm,keepaspectratio=true,angle=0]{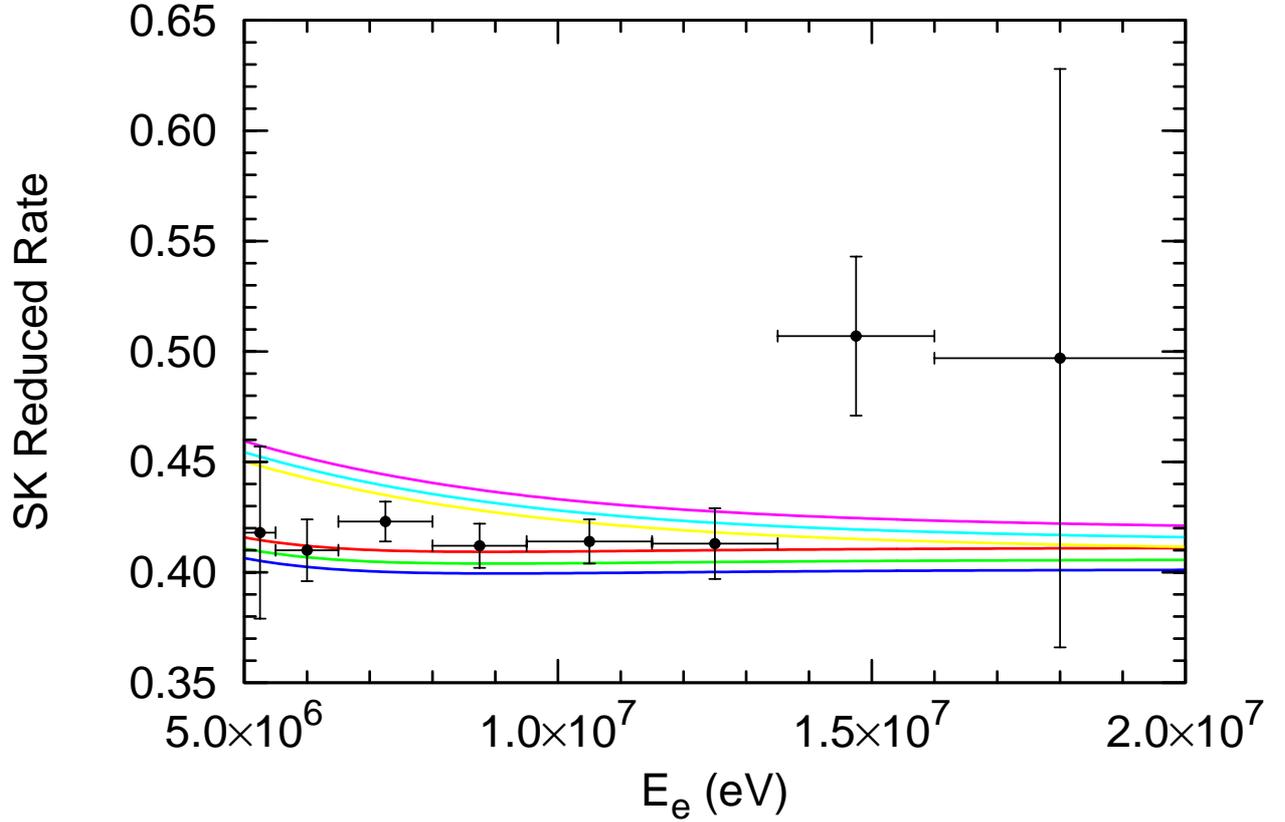}
\caption{ \it The SuperKamiokande spectrum: theoretical predictions and data
points \cite{Fukuda:2002pe} normalized to BPS08(GS) \cite{PenaGaray:2008qe}.
The top three curves refer to $\sin \theta_{13}=0,0.1,0.13$ from top to bottom
in the case of zero magnetic field, and the lower three curves refer to
the same values of $\sin \theta_{13}$ for a sizable field (profile 1), with
$B=140$ kG at the peak. There is a clear preference for a sizable field
possibly related to solar activity, in comparison to a vanishing
one. Units for the observed electron energy $E_e$ are in $eV$.}
\label{fig3}
\end{figure}

\begin{figure}
\centering
\hspace*{-1.5cm}
\includegraphics[height=120mm,keepaspectratio=true,angle=0]{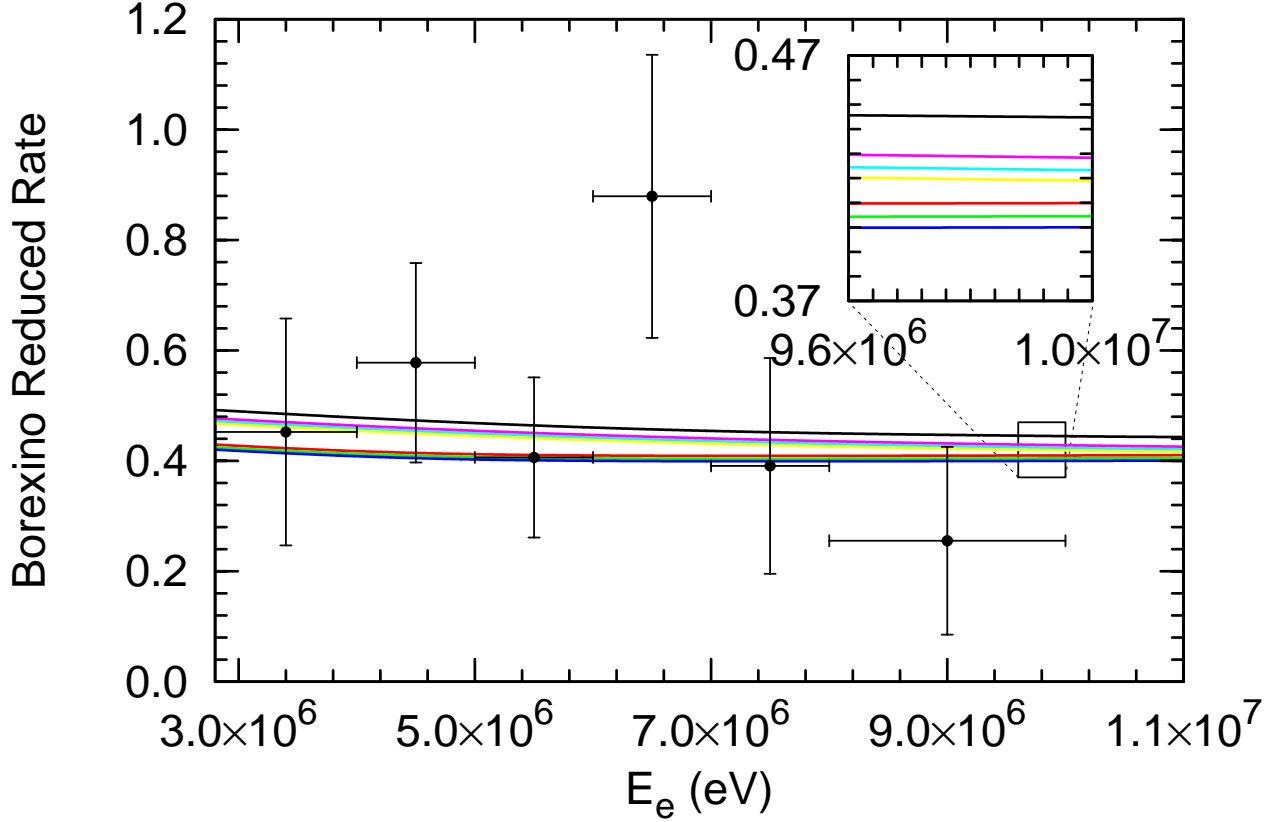}
\caption{ \it The Borexino $^8 B$ spectrum normalized to BPS08(GS)
\cite{PenaGaray:2008qe}. The top curve is extracted from fig.3 of ref.
\cite{Collaboration:2008mr} as explained in the main text. The lower two groups
are from top to bottom the model predictions with $\sin \theta_{13}=0,0.1,0.13$,
$B=0$ and $\sin \theta_{13}=0,0.1,0.13$, $B=140$ kG at the peak (profile 1). There 
is a preference in this case for a vanishing field possibly related to a quiet sun. 
Notice that the theoretical curves coincide for $E_e>5$ MeV with the corresponding 
ones in the previous figure apart from a minor difference due to the energy 
resolution functions.}
\label{fig4}
\end{figure}

\begin{figure}
\centering
\hspace*{-1.5cm}
\includegraphics[height=120mm,keepaspectratio=true,angle=0]{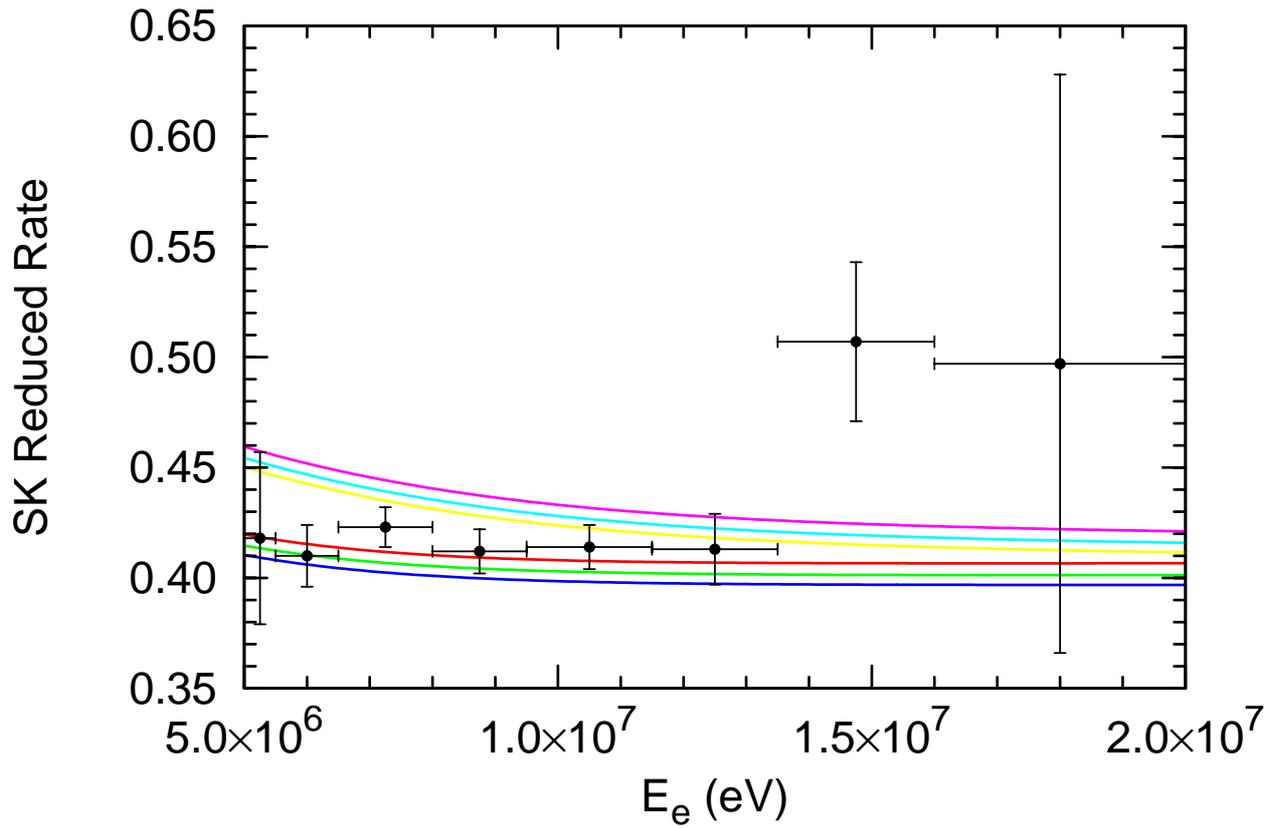}
\caption{ \it Same as fig.3 with the three bottom curves referring to profile 2
with $B=0.75$ MG at the peak.}
\label{fig5}
\end{figure}

\begin{figure}
\centering
\hspace*{-1.5cm}
\includegraphics[height=120mm,keepaspectratio=true,angle=0]{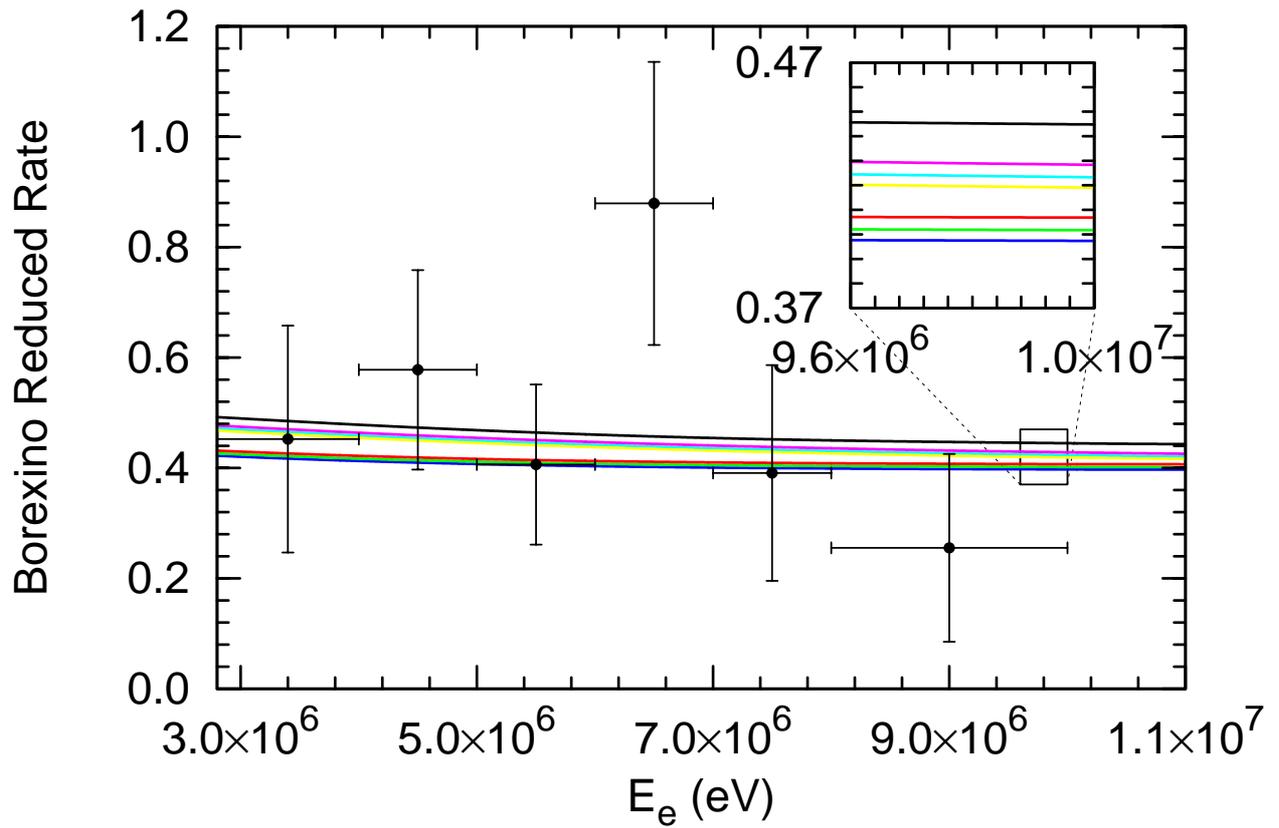}
\caption{ \it Same as fig.4 ($^8 B$ Borexino spectra) for profile 2.}
\label{fig6}
\end{figure}

\begin{figure}
\centering
\hspace*{-1.5cm}
\includegraphics[height=120mm,keepaspectratio=true,angle=0]{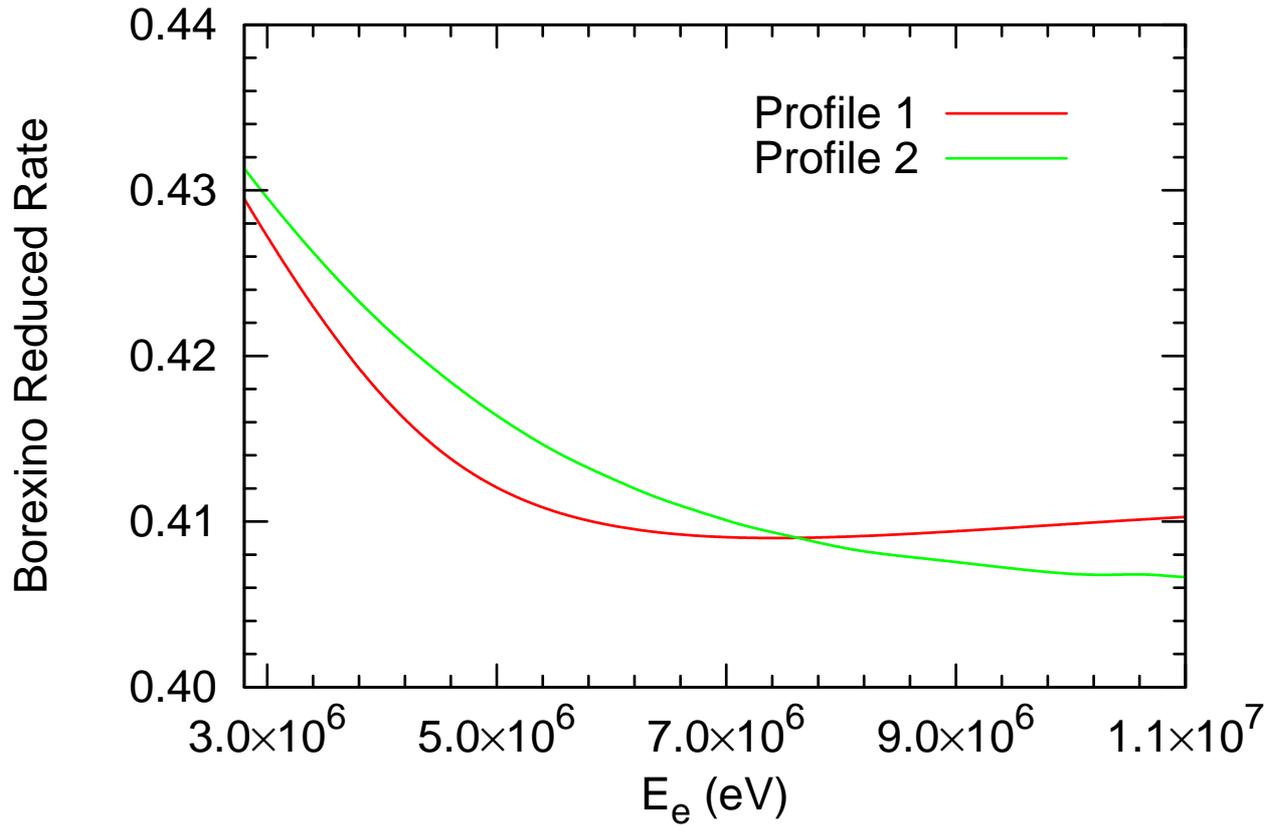}
\caption{ \it Borexino spectra for $^8 B$ neutrinos evaluated for profiles 1
and 2 at the best fit with $\theta_{13}=0$ (parameter values as in the main
text). The spectrum for profile 1 exhibits a shallow minimum while
for profile 2 it is monotonically and smoothly decreasing with the energy.}
\label{fig7}
\end{figure}

\begin{figure}
\centering
\hspace*{-1.5cm}
\includegraphics[height=120mm,keepaspectratio=true,angle=0]{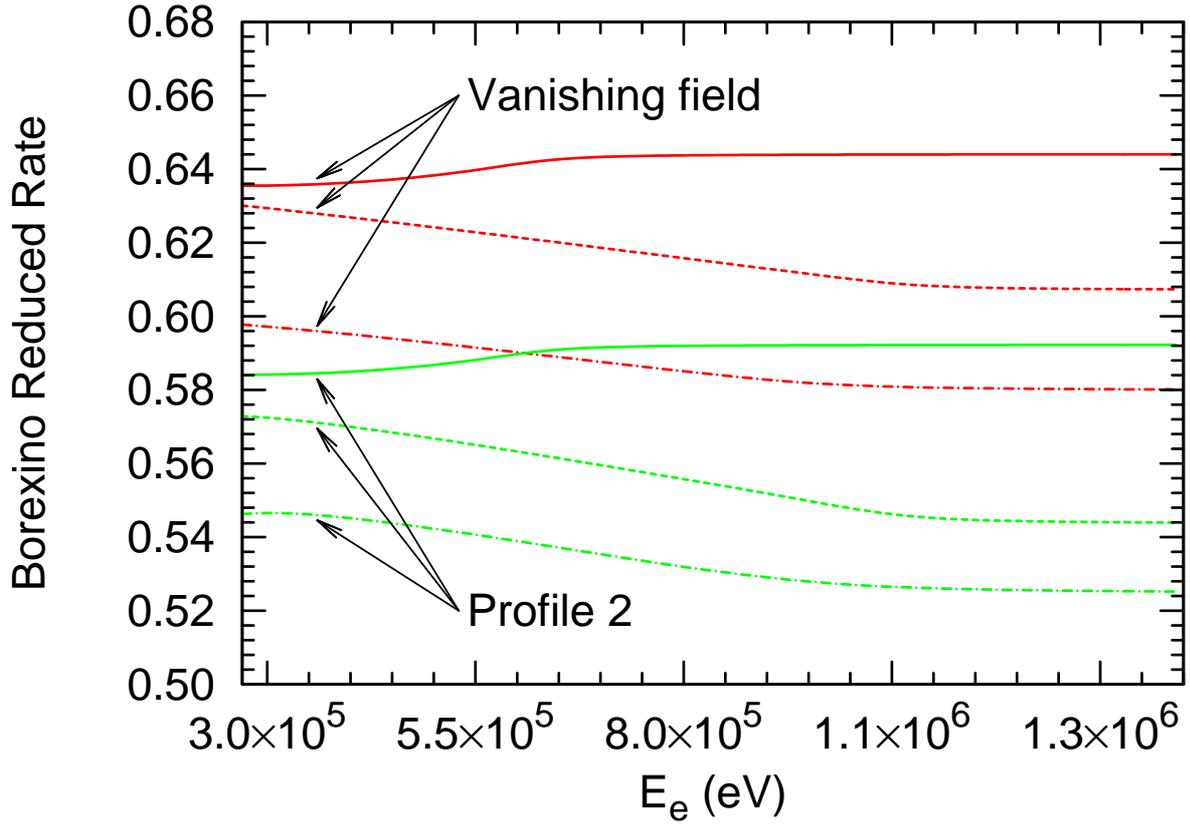}
\caption{ \it Borexino spectra for $^7 Be$ neutrinos (full lines), $^{15} O$
(dashed) and $^{13} N$ (dot-dashed) evaluated for vanishing field and profile 2
at the best fit with $\theta_{13}=0$ (parameter values as in the main
text).}
\label{fig8}
\end{figure}

\end{document}